\documentclass[reprint,prl,aps,groupedaddress]{revtex4-1}

\usepackage{graphicx}
\usepackage{color}
\usepackage{dcolumn}
\usepackage{bm}
\usepackage{amssymb}
\usepackage{color,amsmath}
\usepackage{soul,xcolor} 
\setstcolor{red} 
\usepackage{epstopdf}
\usepackage{booktabs}

\def\wse2{WSe$_2$}
\def\rxx{$R_{xx}$}
\def\rxy{$R_{xy}$}

\newcommand{\comment}[1]{}

\begin{document}

\title{Spin-selective magneto-conductivity in WSe$_2$}

\author{En-Min Shih$^{1}$$^{\alpha}$$^{\beta}$}
\author{Qianhui Shi$^{1}$$^{\gamma}$}
\author{Daniel Rhodes$^{2}$}
\author{Bumho Kim$^{2}$}
\author{Kenji Watanabe$^{3}$}
\author{Takashi Taniguchi$^{4}$}
\author{Kun Yang$^{5}$}
\author{James Hone$^{2}$}
\author{Cory R. Dean$^{1}$}

\affiliation{$^{1}$Department of Physics, Columbia University, New York, NY, USA}
\affiliation{$^{2}$Department of Mechanical Engineering, Columbia University, New York, NY, USA}
\affiliation{$^{3}$Research Center for Electronic and Optical Materials, National Institute for Materials Science, 1-1 Namiki, Tsukuba 305-0044, Japan}
\affiliation{$^{4}$Research Center for Materials Nanoarchitectonics, National Institute for Materials Science,  1-1 Namiki, Tsukuba 305-0044, Japan}
\affiliation{$^{5}$Department of Physics and National High Magnetic Field Laboratory, Florida State University, Tallahassee, Florida 32306, USA
}
\affiliation{$^{\alpha}$Present address: Physical Measurement Laboratory, National Institute of Standards and Technology, Gaithersburg, MD 20899, USA}
\affiliation{$^{\beta}$Present address: Department of Chemistry and Biochemistry, University of Maryland, College Park, MD 10 20742, USA}
\affiliation{$^{\gamma}$Present address: Department of Physics and Astronomy, University of California, Los Angeles, CA, USA}

\maketitle
\textbf{
Material systems that exhibit tunable spin-selective conductivity are key components of spintronic  technologies. Here we demonstrate a novel type of spin-selective transport,  based on the unusual Landau level (LL) sequence observed in bilayer \wse2 under large applied magnetic fields.   We find that the conductivity depends strongly on the relative iso-spin ordering between conducting electrons in a partially filled LL and the localized electrons of lower energy filled LLs, with conductivity observed to be almost completely suppressed when the spin-ratio and field-tuned Coulomb energy exceed a critical threshold. Switching between “on/off” states is achievable through either modulation of the external magnetic or electric fields, with many-body interaction driving a collective switching mechanism.  In contrast to magnetoresistive heterostructures, this system achieves electrically tunable spin filtering within a single material, driven by interaction between free and localized spins residing in energy-separated spin/valley polarized bands. Similar spin-selective conductivity may be realizable in multi-flat band systems at zero magnetic field.}

Spin-dependent transport effects, such as giant magnetoresistance (GMR), play a fundamental role in spintronics. In GMR structures\cite{binasch:1989,fert:2008,telford:2020},   electrons in a non-magnetic conduction layer scatter off moments in the surrounding magnetic layers due to exchange interaction.  The scattering rate is typically enhanced when the conducting electron and the moment have different spins, so that a change in the relative polarization under applied fields causes a large change in resistance\cite{fert:1968}. This effect is exploited in functional devices such as spin valves and magnetic field sensors, and plays an important role in current data storage technologies \cite{chappert:2007}.

Evidence of a homologous effect has been observed in the quantum Hall regime of certain large-mass two-dimensional quantum wells, but remains far less explored. 
In both AlAs and ZnO, a combination of perpendicular and parallel magnetic fields can induce a spin splitting that exceeds the cyclotron gap, $E_{z}/E_{cylc.}>1$. This results in spin-polarized bands, with multiple lower-energy, filled, Landau levels (LLs) polarized with one spin and the LL at the Fermi energy carrying a different spin. It appears that the conducting electrons interact with the localized spins of the filled states such that the longitudinal conductivity is diminished (enhanced) when the Fermi energy is in a spin minority (majority) LL\cite{vakili2005spin,maryenko2015spin}. However, the observed contrast between spin-minority and majority conductivity has been comparatively modest and the precise mechanism has remained unresolved. Nonetheless, these results hint at the possibility of realizing a novel kind of spin-selective transport within a single material with energy-separated spin-polarized bands, rather than a heterostructure with spatially separated magnetic domains.

Here we investigate hole-transport in high mobility bilayer \wse2 in the quantum Hall regime. \wse2 is a van der Waals semiconductor that, like graphene, can be exfoliated to the atomically thin limit. 
Unlike graphene, \wse2 possesses an exceptionally large spin suceptibility\cite{xu2017odd} that leads to an atypical quantum Hall scenario where the Zeeman energy exceeds the cyclotron energy under purely perpendicular fields\cite{fallahazad2016shubnikov,xu2017odd,gustafsson2018ambipolar,shi2020odd,shi2021bilayer}. 
Together with spin-valley locking, this gives rise again to a highly polarized LL sequence, where a large number of the lowest energy bands all share the same iso-spin polarization \cite{gustafsson2018ambipolar,shi2020odd,shi2021bilayer} (Fig. 1a, b). We find that \wse2 exhibits a similar spin-dependent conductance as in polarized AlAs and ZnO, but with a much more dramatic effect.  The conductance within partially filled majority- and minority-spin LLs shows opposite temperature dependence, with the minority-spin LL conductance approaching zero at low temperature and high magnetic field.  This observation identifies a regime of essentially perfect spin selectivity arising from a complete localization behaviour for the minority carriers.

\begin{figure*}[ht]
\includegraphics[width=\linewidth]{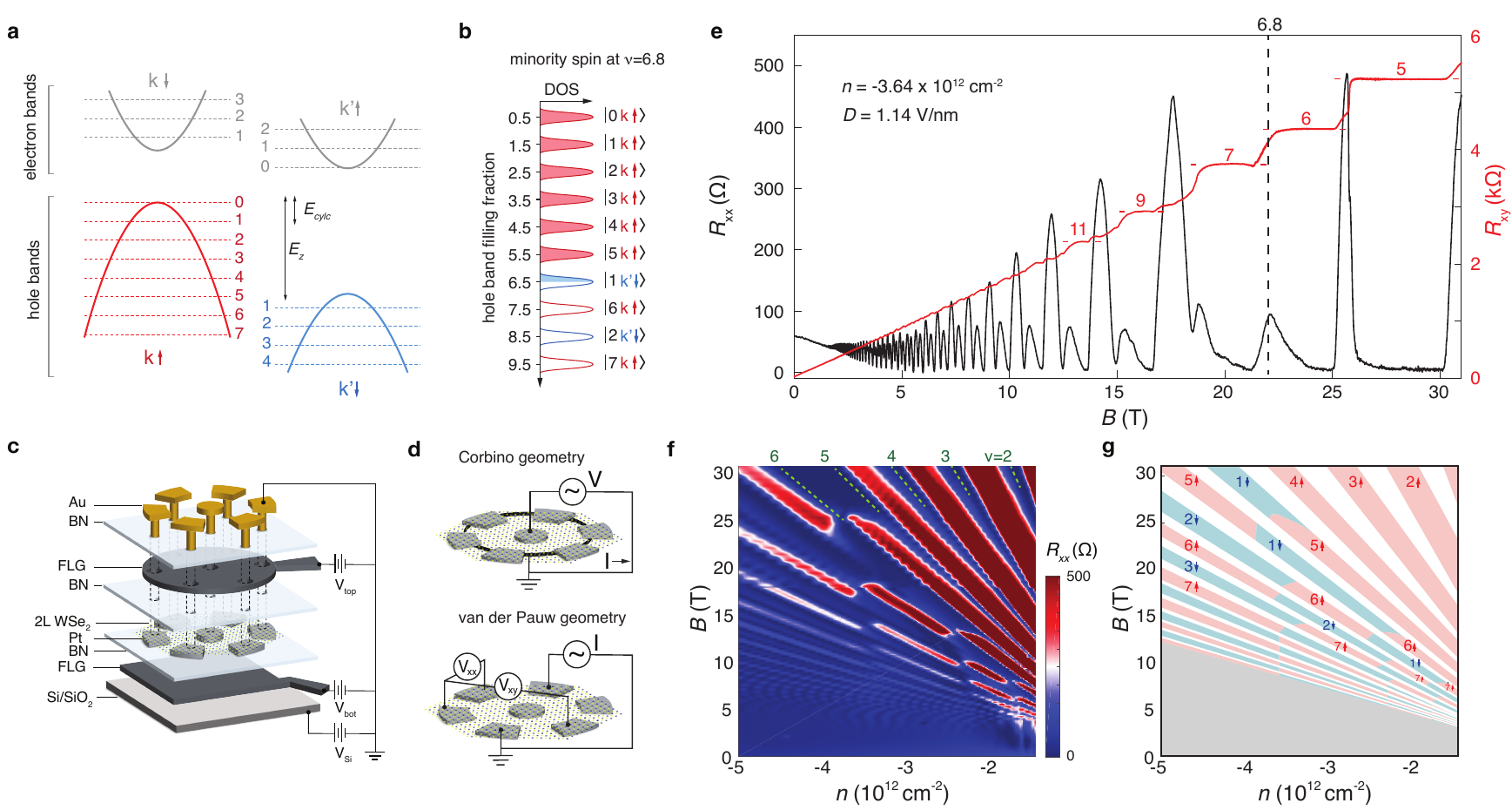} 
\vspace{-0.15 in}
\centering
\caption{
\textbf{Device geometry and spin-dependent magneto-transport.}
(a) Schematic of single layer \wse2 bands under magnetic field. The Landau level (LL) sequences in different valley and carrier bands are labeled.
(b) Hole filling of LLs at filling fraction $\nu = 6.8$, at which the first minority spin LL is being filled.
(c) Schematic of the dual-gated bilayer \wse2 device.
(d) Two measurement geometries. Top panel: two-probe measurement in Corbino configuration. Bottom panel: four-probe measurement in van der Pauw configuration.
(e) R$_{xx}$ and R$_{xy}$ versus magnetic field in van der Pauw configuration at $n = -3.64 \times 10^{12}$ cm$^{-2}$ with only top layer of bilayer \wse2 populated.
(f) Mapping of R$_{xx}$ versus magnetic field and density. The first few LL filling factors are labeled. 
(g) Schematic of (f) with the orbital number and spin state of valence LL labeled.
}
\label{fig:1}
\vspace{-0.15 in}
\end{figure*}

\begin{figure*}[ht!]
\includegraphics[width=6.5 in]{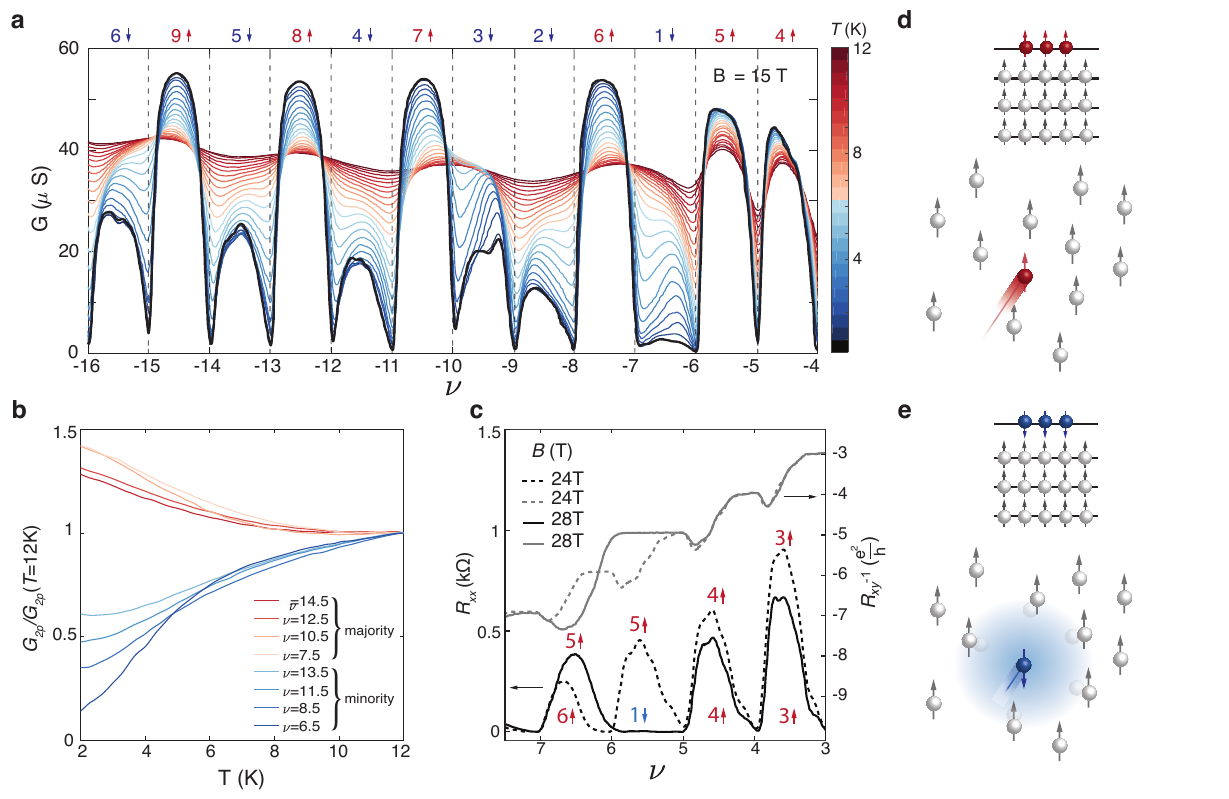} 
\vspace{-0.15 in}
\caption{\textbf{Electronic transport signatures of exotic localization in minority spin LL.}
(a) Two-probe conductance versus filling factor in Corbino configuration at B = 15 T at different temperature. The base temperature (black line) is 0.8 K. 
(b) Normalized two-probe conductance versus temperature at different half-fillings at B = 15T. The conductance is normalized with conductance at 12K.
(c) \rxx and \rxy versus filling factor at B = 24 T and B = 28 T. At $\nu=5.5$, the valence LL is in spin-majority state and spin-minority state at B = 24 T and B = 28T, respectively.
(d)(e) Cartoon illustrations of the spin-dependent transport mechanism. In (d) the valence LL is in spin-majority state and the mobile charge can flow through the immobile same-spin charge background freely, while in (e) the valence LL is in spin-minority state and the mobile charge is auto-localized due to dressing the distorted immobile opposite spin charges.  
}
\label{fig:2}
\vspace{-0.15 in}
\end{figure*}

We suggest that the spin dependent localization may be understood from a polaron-like model, in which  a minority spin carrier displaces the background incompressible majority spin carriers and becomes self-localized.  By studying this effect in blayer WSe2, we show that the spin-dependent transport can be generalized to real spin or valley pseudospin. Finally, we demonstrate that switching between the high/low conductance states can be initiated within a single, partially filled, LL when Coulomb interaction drives a collective spin flip. These results provide a prototypical demonstration of electric-field tunable spin-dependent transport in an all-electron system, where Coulomb interactions and exchange effects play central roles.

We measured magneto-transport in dual-gated bilayer \wse2 fabricated with pre-patterned Pt contact elelctrodes\cite{fallahazad2016shubnikov} (Fig. 1c).
The bilayer \wse2 was exfoliated from a flux-grown single crystal, with charged defect density lower than $10^{11}$ cm$^{-2}$\cite{edelberg2019approaching}.
Using the van der Waals stacking approach\cite{Lei2013edgecontact} together with standard nanofabrication technqiues, we implemented a novel device geometry that supports measurements of both edge and bulk transport in the same device (Fig. 1d).  Six contacts arranged in a circular pattern enable four-terminal longitudinal ($R_{xx}$) and Hall resistance ($R_{xy}$) measurements in a``van der Pauw geometry''(Fig. 1d bottom).  Additionally, a center contact enables measurement of the two-terminal conductance between the center and edge contacts probes the bulk response in the Corbino geometry (Fig. 1d top). Dual gates provide independent control of the carrier density of the two layers \cite{shi2021bilayer,zeng2013optical,xu2014spin,fallahazad2016shubnikov,pisoni2019absence}, allowing us to study the device response with either a single layer or both layers populated. We restrict our measurement here to the hole band response only.

We first focus on the regime with only one layer populated, where the device effectively behaves like an isolated monolayer. Figure 1e shows $R_{xx}$ and $R_{xy}$, measured in the van der Pauw geometry as a function of applied perpendicular field $B$, with hole density tuned to n$_\mathrm{hole}=3.64\times10^{12}$ cm$^{-2}$, and transverse displacement field, $D=1.14$~$V/$nm. The Hall mobility, measured in the low field regime, is $\mu_{H} \approx 30,000$ cm$^2$/Vs, which is the highest reported for any semiconductor TMD\cite{lin2019determining, movva2015high}. 
Quantum oscillations appear below 2 T, giving an approximate estimate for the Landau level broadening of 5.92 K, which is less than has been reported for high mobility graphene\cite{zeng2019high}. In the high field regime, fully quantized Hall plateaus concomitant with zero longitudinal resistance are observed at integer fillings, further confirming the high quality of the device.

The most dramatic feature seen in the magnetoresistance (Fig. 1e) is a an alternating high/low  longitudinal resistivity between neighbouring LLs. Previous studies of \wse2 established the LL spin splitting to be density- and field-dependent, giving rise to the approximate ladder of states shown in Fig. 1a\cite{gustafsson2018ambipolar,shi2020odd}. At magnetic fields on the order 15~T, the lowest 6 LLs are fully iso-spin polarized, with higher fillings corresponding to alternating majority/minority iso-spin. We find that the resistance modulation correlates with the relative spin order at the Fermi energy with high (low) $R_{xx}$ coinciding with the spin majority (minority) LLs. For example, the peak resistance at $\nu=6.8$, which corresponds to 6 filled polarized levels and one partial filled anti-poplarized (Fig. 1b), has resistance less than 1/5 the value at partial filling of neighbouring, spin-majority, LLs. The general influence of the spin state on transport is unambiguously demonstrated in the plot of $R_{xx}$ versus field and density shown in Fig. 1f. The $R_{xx}$ value periodically switches between high value (red) and low value (blue), matching the spin transition map shown in Fig. 1g. 

In the high-field, low-density regime,  the longitudinal conductivity, $\sigma_{xx}$, is approximately proportional to the longitidunal resistivity, $\rho_{xx}$, such that low resistance indicates low conductance. The Corbino geometry is used to probe the bulk conductance directly, i.e. independent of any potential edge effects.  Figure 2a shows the two-terminal conductance in the Corbino geometry,  acquired at $B = 15$T and varying temperature. 
At T = 0.8K (black line), majority-spin LLs exhibit high conductance and minority spins exhibit low conductance, consistent with the van der Pauw resistivity measurements. The temperature dependence likewise shows two behaviours (Fig. 2b).  For majority-spin fillings, the conductance exhibits metallic-like response, \textit{i.e.} increases with decreasing temperature, as is typical for normal QH systems at partial filling.  For the minority spins, the conductance decreases with decreasing temperature, indicating an insulating-like bulk response consistent with localization of the carriers.

The localization strength varies with both total filling fraction and magnetic field. In the Corbino measurement at $B=15$~T (Fig. 2a), the spin-minority conductance is minimal at filling fraction 6.5, corresponding to a single anti-polarized LL in a background of fullly polarized LLs, and then increases monotonically with increasing LL index. Figure 2c shows the effect of varying the magnetic field, measured in the van der Pauw geometry. As the magnetic field is increased from 24 T, filling fraction at 5.5 undergoes a transition from spin majority to spin minority, evidenced by transition from large \rxx (dashed black line) to small \rxx (solid black line). By around 28 T, \rxx completely vanishes across the full LL, within measurement resolution. Simultaneously, the $\nu=5$ Hall plateau extends across this same filling range.  This remarkable observation suggests that the spin-minority LL is fully localized within the entire LL.

\begin{figure*}[ht!]
\includegraphics[width=5.5in]{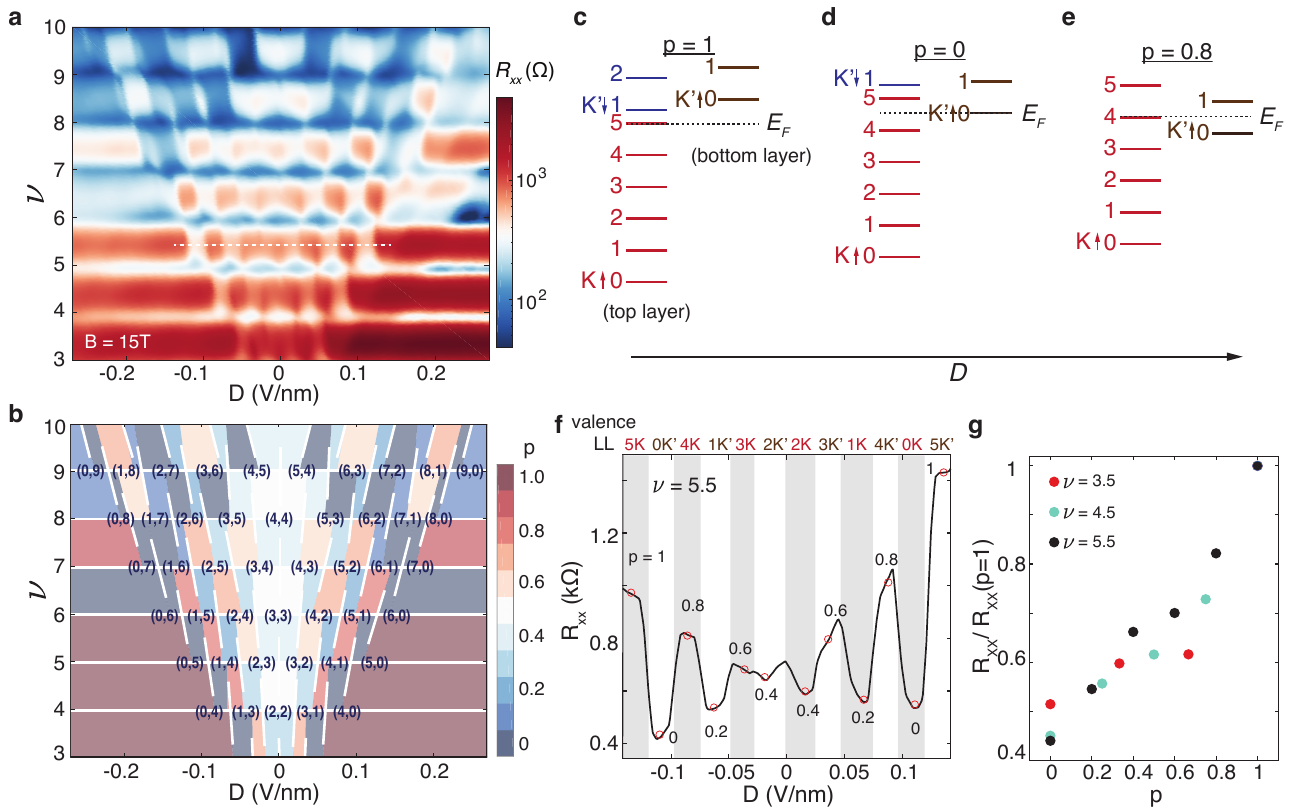} 
\vspace{-0.15 in}
\caption{\textbf{Flavor-dependent transport in the bilayer region.}
(a) Mapping of R$_{xx}$ versus total filling factor and displacement field.
(b) Schematic diagram of (a) with filling factors of the two layer labeled. At partial filling, the state is color coded with the p value. 
(c)(d)(e) Examples of p value at different LL configurations. (c) The valence LL species occupies all the filled LLs, $p=1$. (d) No valence LL species in the filled LLs, $p=0$. (e) Four valence LL species among total five filled LLs, $p=0.8$.
(f) \rxx plotted with displacement field at $\nu_{tot}=5.5$. At half-filling of every valence LL, the p value and the specie of valence LL are labeled.
(g) Normalzed \rxx versus p value at different $\nu_{tot}$. The \rxx is normalied to the value at $p=1$.
}
\label{fig:3}
\vspace{-0.15 in}
\end{figure*}

\begin{figure*}[ht!]
\includegraphics[width=4.5in]{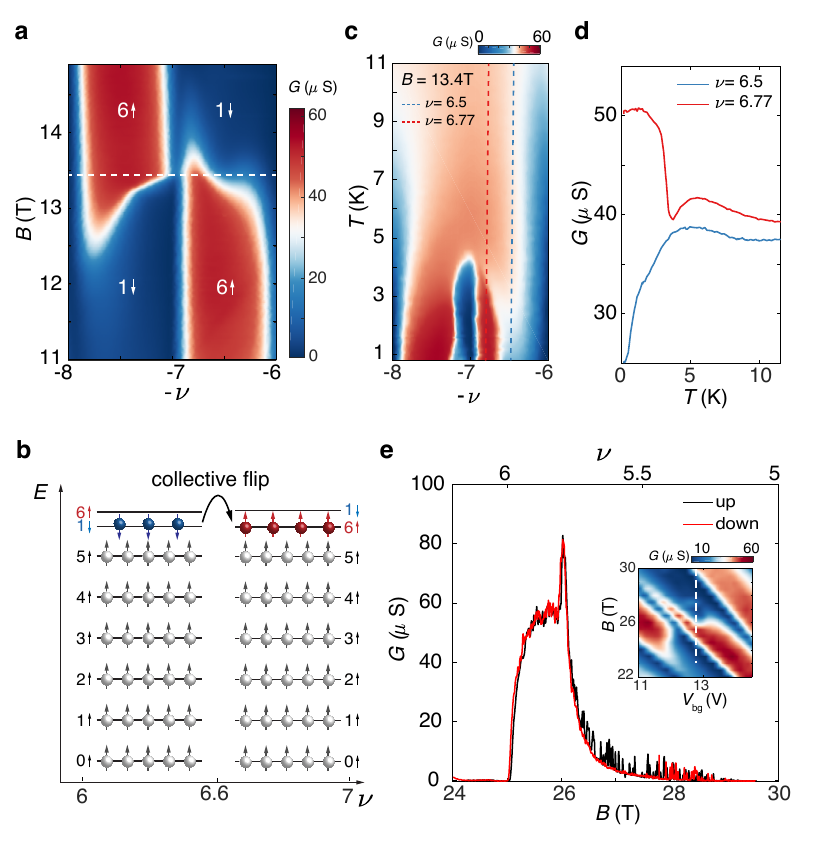} 
\caption{\textbf{Minority-spin to majority-spin transition within valence LL.}
(a) Mapping of conductance versus magnetic field and filling factor around the transition.
(b) Cartoon illustration of the transition in (a). 
(c) Mapping of conductance versus temperature and filling factor along B = 13.4 T (white dashed line in (a)).
(d) The linecuts of conductance versus temperature along $\nu=6.5$ (red dashed line in (c)) and $\nu=6.77$ (blue dashed line in (c)).
(e) Conductance versus magnetic field (or filling fraction) along the the white dashed line shown in the inset. Inset: mapping of conductance versus magnetic field and back-gate voltage.
}
\label{fig:4}
\end{figure*}

Both the filling fraction and magnetic field dependence of the minority carrier localization can be understood from the spin-dependent exchange interaction between the mobile and localized carriers.  Mobile carriers interact weakly (strongly) with localized carriers that share the same (opposite) iso-spin polarization. The effect on transport is illustrated schematically in Fig. 2d, e.  Mobile carriers in a partially-filled spin-majority LL scatter only weakly from the localized carriers (Fig. 2d). For spin-minority carriers, the strong Coulomb interaction can distort the immobile background charges, such that the free carrier becomes effectively more massive and less conductive (Fig. 2e).  
This is analogous to a polaron, but in an all electron system with the immobile carriers in the filled LLs playing the equivalent role of a crystal lattice.  Since in the QHE regime the Coulomb energy scales with the magnetic field, $E_c=e^2/\epsilon l_B\propto\sqrt{B_{\perp}}$, the minority-carrier should become increasingly localized with increasing field, consistent with our observation.  On the other hand, as filling fraction is increased at fixed magnetic field, the net polarization of the carriers reduces towards zero, reducing the difference between majority and minority spin mobilities, also consistent with observation. The picture identified here may also explain why the observed effect is magnified in \wse2 compared with previous studies. The influence of the filled LLs can be characterized by the LL mixing parameter\cite{sodemann2013landau,peterson2013More,simon2013Landau}, $\kappa=E_{Coulomb}/E_{cycl.}$, i.e. the ratio of the Coulomb energy, to the cyclotron gap.  For comparison we note that $\kappa=\alpha/\sqrt{B}$, with $\alpha=2.6, 16.7, 22.5, 31$ in GaAs, ZnO, AlAs and WSe2, respectively. The unusually large $\kappa$,  together with the large ratio of $E_z/E_{cycl.}$  creates a unique scenario in \wse2 with simultaneously larger polarization and larger mixing than in previously studied systems.

\comment{
\textcolor{red}{The screening capability depends on the partial filling factor in this strongly-interacting system.

The polarizability $\Pi$ of carriers in the filled LLs peaks near a momentum q$\approx1/l$, where $l$ is the magnetic length\cite{aleiner1995two}. 
Therefore, the interaction between minority carriers and the backround majority carriers is optimally screened when the minority-spin charges are localized at a length scale $l$. As the filling of minority spin carriers increases, their wavefunctions overlap and the potential from them becomes more uniform and no longer localizes around $l$. Therefore, to minimize the interaction energy, the minority-spin carriers further localize themselves by mixing with higher LLs. 
The LL mixing in turn pushes the extended states of the valence LL higher in energy \cite{haldane1997landau}; this process stops only when the majority spin electrons start to enter the valence LL. This phenomenon can explain our observation in Fig.2(c) that conduction due to extended states is missing in a large range of filling factor $5 < \nu <6$, with \rxx remaining at zero and \rxy at the plateau value.[I don't really understand this text]}
}

Both the real spin and the valley pseudospin determine the exchange interaction. In bilayer \wse2, the strong spin-orbit coupling and the stacking order result in coupled spin, valley and layer degrees of freedom. A total of four flavors are relevant: $(\uparrow, K)$, $(\downarrow, K')$ localized in the top layer, and $(\uparrow,K')$, $(\downarrow,K)$ localized in the bottom layer \cite{shi2021bilayer}.
Therefore, the valley degree of freedom can be controlled by a displacement field, $D$, which moves carriers from one layer to another. 
In Fig. 3a we plot \rxx measured at 15 T versus $D=(V_{T}-V_{B})/2\epsilon_{0}$ and the total filling factor $\nu_{tot}=hC(V_{T}+V_{B})/e^2B$, where $V_{T}$ and $V_{B}$ are top gate and bottom gate voltage bias, and $C$ is the geometric capacitance.

The resistance shows a checkerboard-like pattern as charge is transferred between different LLs localized in the two layers \cite{shi2021bilayer}. 
For $\nu < 6$, all LLs have the same spin, but charges in the top layer reside in the K valley and those in the bottom layer are in the K' valley.
  
Examples of the change in LL configuration versus $D$ for fixed filling fraction, $\nu=5.5$, are shown in Fig. 3c-e. At large negative $D$, the free and localized electrons share the same spin and valley.   As $|D|$ is decreased from the fully valley-polarized regime, the Fermi level switches between valley K and K'.  

Figure 3f shows a plot of \rxx versus $D$ at $\nu = 5.5$, corresponding to a linecut along the dashed line in Fig. 3a.  \rxx oscillates as the Fermi level switches between majority (large \rxx) and minority (small \rxx) valley spin. The resistance is approximately symmetric about $D=0$, confirming that the value of \rxx is dominated by the relative net iso-spin order, rather than other features such as the spin or valley flavour. To quantify how \rxx varies with the relative iso-spins we introduce the parameter, 
$p=N_{F}/N_{tot}$, where $N_{tot}$ is the total number of filled LLs, and $N_{F}$ is the number of filled LLs that have the same flavor as the LL at the Fermi energy.  We label the peaks and minima in Fig. 3f by these $p$ values and see that indeed \rxx is maximized (minimized) when $p$ is large (small). 

Figure 3g plots the evolution of the  normalized longitudinal resistance $R_{xx}/R_{xx}(p=1)$ versus $p$ for three  filling factors 3.5, 4.5 and 5.5.  $R_{xx}/R_{xx}(p=1)$ shows an approximately linear increase with $p$ confirming the longitudinal resistance is proportional to the net population difference between the free carrier flavor and the localized carrier flavors. In Fig. 3b, we colour distinct regions in the $D -\nu$ phase space by the magnitude of $p$.
For $\nu > 6$, $p$ is calculated as the population of the valence flavor among all four flavors.
The consistency of the polarization color map with the \rxx color map in Fig. 3a again indicates correlation between $p$ and the resistance.
These observations are consistent with our picture that localization of the mobile carriers at the Fermi energy is driven by interaction with incompressible carriers of a different flavor: the greater the population of different-flavored incompressible carriers, the lower the conductance.

Finally, we discuss the switching between  majority/minority flavors within a single partially filled LL. Figure 4a plots the conductance versus magnetic field and filling factor around an iso-spin transition due to a LL crossing. As before, high (low) conductance correlates with iso-spin majority (minority) free carriers.  In detail we see that the conductance transition from high to low value onsets at partial fillings across a well defined and narrow boundary. This suggests a collective transition that reorders the spin at the Fermi level (Fig. 4b)\cite{hunt2017direct,shkolnikov2005observation,pisoni2018interactions}.  Figure 4c plots the conductance at $B$ = 13.4 T versus the filling factor (white dashed-line in Fig. 4a) and temperature.  The temperature dependence of the conductance at two filling factors ($\nu = 6.55$ and $6.77$) are compared in Fig. 4d. The temperature dependence is effectively identical down to approximately 5~K, below which the trends suddenly diverges with $\nu = 6.55$ dropping sharply towards zero (consistent with minority carrier localization) while $\nu = 6.77$ shows metallic like response. The sudden change in beahviour below an apparent critical temperature further supports the notion of a many-body driven spin-transition when the temperature drops below a characteristic interaction energy scale. 

The collective flip of many spins with increasing filling factor is evocative of LL levitation driven by Coulomb interactions\cite{haldane1997landau}. With increasing partial filling factor in a minority-spin LL, the strong Coulomb interaction between the minority-spin carriers at the Fermi level and the majority-spin carriers in the background (and the urge to optimally screen this interactions) effectively pushes the valence LL up. This LL levitation with increasing partial filling explains the missing extended-states at high magnetic field (Fig. 2c).

Finally, when the filling factor is tuned by sweeping the magnetic field at a fixed density, a large conductance spike is observed at the transition. This typically suggests the formation of domain walls \cite{jungwirth2001resistance} between the two different spin states indicating a first-order transition (Fig. 4e).

In summary, we have shown strong spin-dependent transport in bilayer WSe$_2$. We demonstrate that spin/valley-polarized band arrangement in energy space can give strong spin/valley-selective transport, achieving perfect selectivity in the extreme limit. The ability to engineer polarized flat bands in van der Waals heterstructures, such as through moire patterning, might enable a similar electric-field driven iso-spin selectivity under zero magnetic fields.  This type of bandstructure driven GMR could pave the way for next-generation spintronic devices.

We thank Luis Balicas, William Coniglio and Bobby Pullum for help with experiments at the National High Magnetic Field Lab. This research is primarily supported by US Department of Energy (DE-SC0016703). Synthesis of \wse2 (D.R.,B.K.,K.B.) was supported by the Columbia University Materials Science and Engineering Research Center (MRSEC), through NSF grants DMR-1420634 and DMR-2011738. The work of KY was supported by the National Science Foundation Grant No. DMR-1932796. A portion of this work was performed at the National High Magnetic Field Laboratory, which is supported by National Science Foundation Cooperative Agreement No. DMR-1644779 and the State of Florida. K.W. and T.T. acknowledge support from the JSPS KAKENHI (Grant Numbers 20H00354, 21H05233 and 23H02052) and World Premier International Research Center Initiative (WPI), MEXT, Japan.

\bibliography{main}

\newpage

\newpage
\clearpage

\pagebreak
\begin{widetext}
\section{Supplementary Materials}

\begin{center}
\textbf{\large Spin-selective magneto-conductivity in WSe$_2$}\\
\vspace{10pt}
E. Shih, Q. Shi, D. Rhodes, B. Kim, K. Watanabe, T. Taniguchi, K. Yang, J. Hone, C. R. Dean\\ 
\vspace{10pt}
\end{center}

\noindent\textbf{This PDF file includes:}\\
\noindent{Supplementary Text}\\
\noindent{Materials and Methods}\\

\newpage
\renewcommand{\thefigure}{S\arabic{figure}}
\renewcommand{\theequation}{S\arabic{equation}}
\renewcommand{\thetable}{S\arabic{table}}
\setcounter{figure}{0} 
\setcounter{equation}{0}

\section{S1. Device Fabrication and Geometry}

The cross-section of bilayer WSe$_2$ device described in the main text is shown in Fig. S1a. The full stack is made of two pieces: the bottom stack (Fig. S1a blue dashed block and photo in Fig. S1c) is made of BN encapsulated graphite followed by deposition of prepatterned contact Cr/Pt (2nm/20nm), and the top stack (Fig. S1a red dashed block and photo in Fig. S1b) is made by picking layers in order of BN/Graphite/BN/2L WSe$_2$. The top stack is then put on the bottom stack with 2L WSe$_2$ contacted with the Pt (Fig. S1d). Next, the partial etching step shapes the top-gate which defined the channel area (Fig. S1e). Then, another BN is put on top of stack to avoid shorting the top-gate to the later evaporated leads (Fig. S1f). In this design, leads can be placed across the channel and make Corbino geometry possible. Finally, holes are etched to the Pt prepattern (Fig. S1g) followed by deposition of channel leads and contact-gate (Fig. S1h). The device is designed so that the contact-gate can turn on the contact region (with large negative voltage), and top/bottom gate voltage can be positive or negative; therefore the full density and displacement field phase space can be mapped out. In device a (Fig. S1h), because the contact gate short all the outer leads together, we have to use negative topgate voltage to turn on contact region, which leaves top-layer always on. In device b (Fig. S2), the contact channel doesn't short the leads; therefore, we can map out the full phase space as shown in main text Fig. 3. 
\vspace{-10pt}
\begin{figure}[hb]
\begin{center}
\includegraphics[width=5in]{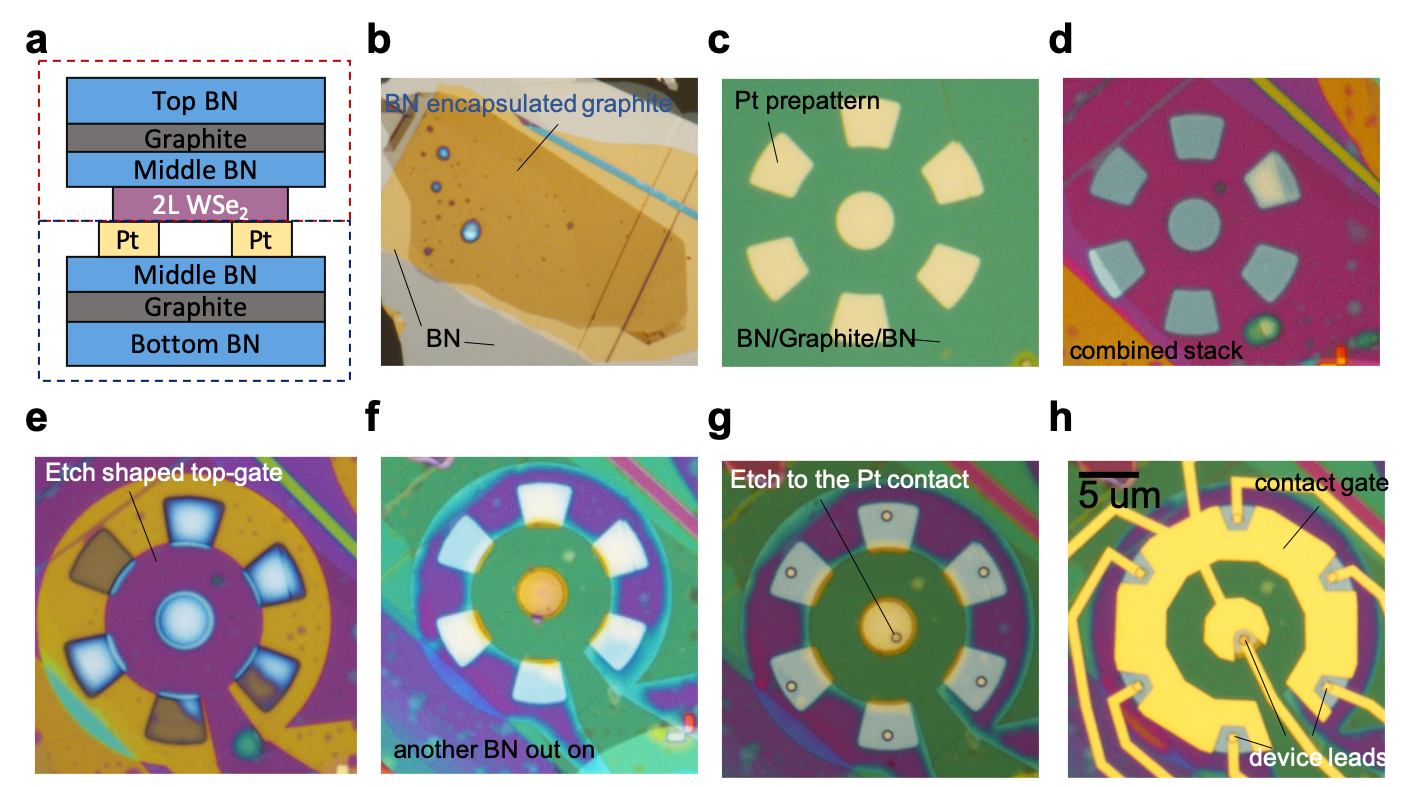}
\vspace{-0.2 in}
\end{center}
\caption{
	\textbf{a,} Full stack schematic. Optical image of
	\textbf{b,} Top stack,
        \textbf{c,} Bottom stack,
	\textbf{d,} Combined top and bottom stack,
	\textbf{e,} Etch-shaped top gate,
	\textbf{f,} Another BN put on,
	\textbf{g,} Etching to the Pt prepattern
	\textbf{h,} Evaporated electrodes of contact-gate and channel leads. Finished device a.
}
\vspace{-0 in}
\end{figure}

\begin{figure}[hb]
\begin{center}
\includegraphics[width=3.5in]{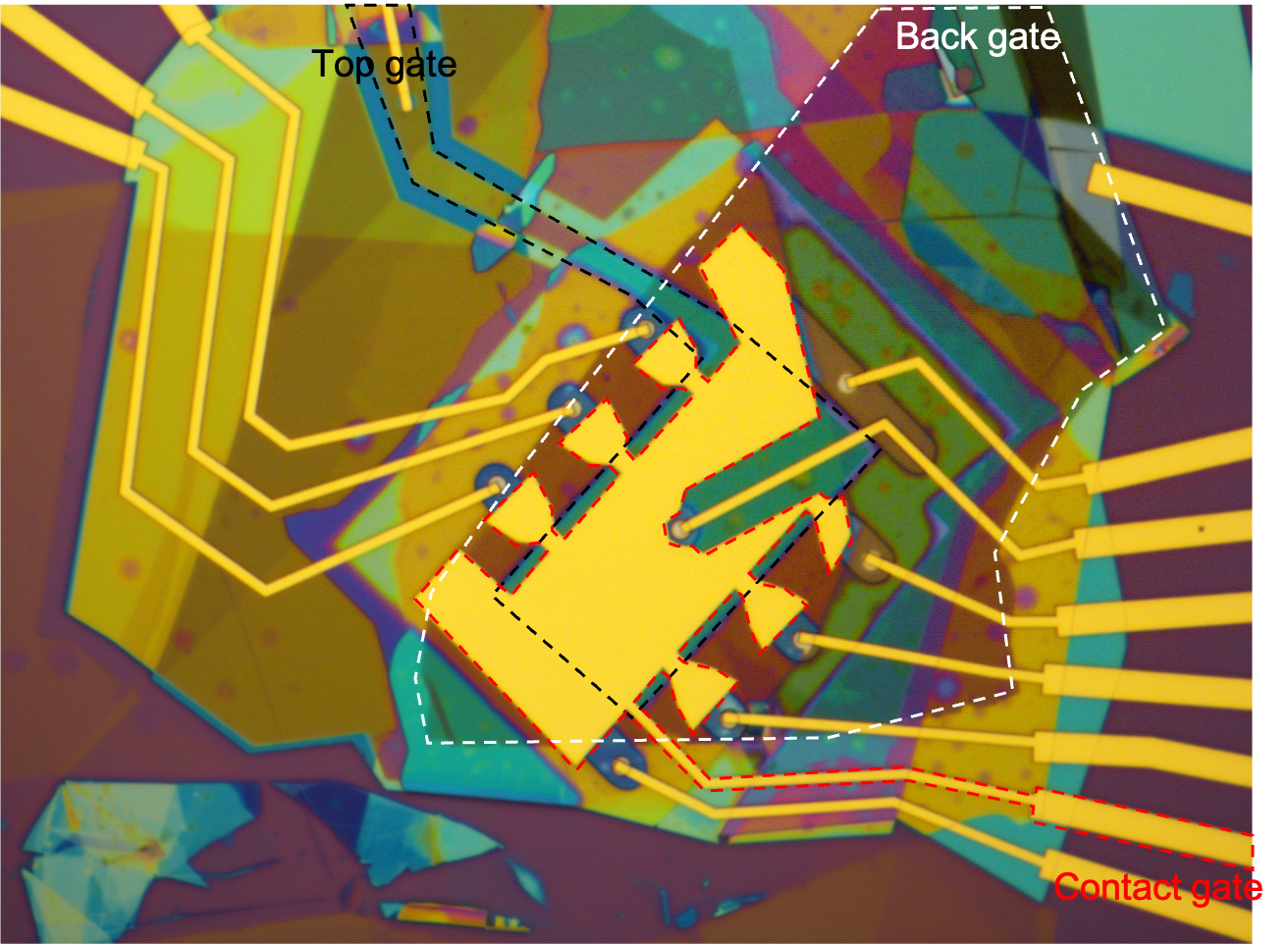}
\vspace{-0.2 in}
\end{center}
\caption{
        Device b with three (top, bottom, contact) gates labeled.
}
\vspace{-0 in}
\end{figure}
\newpage
\section{S2. Landau Level Structure in Bilayer WSe$_2$}
The four iso-spin flavors of bilayer \wse2 are protected by spin-orbit coupling and inversion symmetry of the crystal (more discussion see SI of Ref. \cite{shi2021bilayer}). Under magnetic field, each flavor band split into Landau levels and their relative ordering depends on Zeeman energy (E$_z$) and electric field energy across the layers (E$_L$), as schematically shown in Fig. S3.

\begin{figure}[hb]
\begin{center}
\includegraphics[width=5.5in]{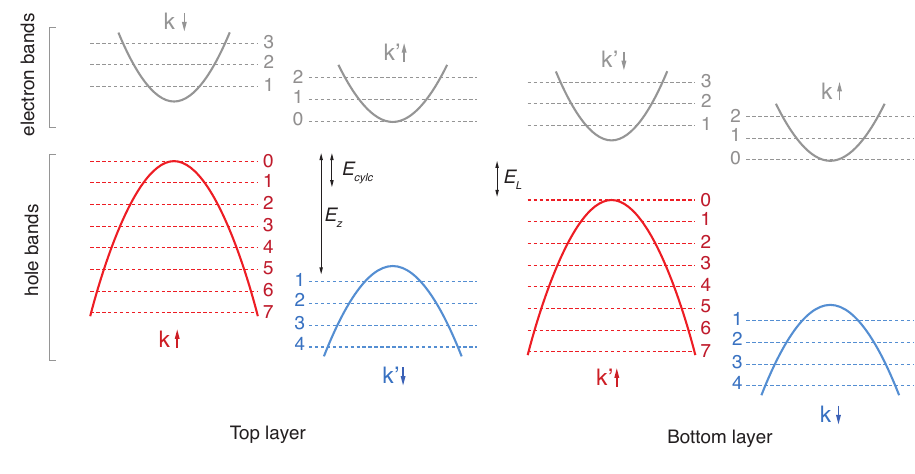}
\vspace{-0.2 in}
\end{center}
\caption{
        Schematic of bilayer \wse2 bands under magnetic field. The Landau levels of different iso-spin flavors are labeled.
}
\label{}
\vspace{-0 in}
\end{figure}

\section{S3. Comparing Different 2D Charge Carrier Systems}

\begin{table}[hb]
\centering
\begin{tabular}{c|cccccc}
\toprule[1.5pt]
 & GaAs (electron) & GaAs (hole) & AlAs & ZnO & Graphene & WSe2 \\
\midrule[1pt]
$m^*/m_0$ & 0.069 & 0.38 & 0.46 & 0.29 &  & 0.45 \\ \hline
$\epsilon/\epsilon_0 $ & 13 & 13 & 10 & 8.5 & 6-7 & 7 \\ \hline
$g^*$ & 0.44 &  & 4-6 & 4 & 4-8  & 9-20 \\ \hline
$E_c(=e^2/\epsilon l_B$) (K) & 50 $\sqrt{B_\perp}$ & 50 $\sqrt{B_\perp}$& 64.7 $\sqrt{B_\perp}$& 75 $\sqrt{B_\perp}$& 108$\sqrt{B_\perp}$ & 93 $\sqrt{B_\perp}$\\ \hline
$E_{cycl}(=\hbar \omega)$ (K) & 19.4 B & 3.5 B & 2.9 B & 4.6 B &  216 B & 2.96 B \\ \hline
$\kappa (=E_c/E_{cycl})$ & 2.6/$\sqrt{B_\perp}$& 14.6/$\sqrt{B_\perp}$& 22.5/$\sqrt{B_\perp}$& 16.6/$\sqrt{B_\perp}$ & 0.5-2.2 & 31/$\sqrt{B_\perp}$ \\ \hline
$ E_z/E_c (=g^*m^*) $ & 0.03 &  & 1.84-2.76 & 1.16 & 0.01-0.04 & 4-9 \\ \bottomrule[1.5pt]
\end{tabular}
\caption{Comparison of characteristics of 2D charge carrier systems, including GaAs, AlAs\cite{sodemann2013landau, padmanabhan2010composite}, ZnO\cite{sodemann2013landau, maryenko2012temperature}, graphene\cite{young2012spin}, and \wse2\cite{fallahazad2016shubnikov}(B is measured in Tesla).}
\end{table}
\newpage

\section{S4. Landau Level Levitation}
The screening capability depends on the partial filling factor in this strongly-interacting system.
The polarizability $\Pi$ of carriers in the filled LLs peaks near a momentum q$\approx1/l$, where $l$ is the magnetic length\cite{aleiner1995two}. 
Therefore, the interaction between minority carriers and the backround majority carriers is optimally screened when the minority-spin charges are localized at a length scale $l$. As the filling of minority spin carriers increases, their wavefunctions overlap and the potential from them becomes more uniform and no longer localizes around $l$. Therefore, to minimize the interaction energy, the minority-spin carriers further localize themselves by mixing with higher LLs. 
The LL mixing in turn pushes the extended states of the valence LL higher in energy \cite{haldane1997landau}; this process stops only when the majority spin electrons start to enter the valence LL. This phenomenon can explain our observation in main text Fig.2(c) that conduction due to extended states is missing in a large range of filling factor $5 < \nu <6$, with \rxx  remaining at zero and \rxy at the plateau value.

\section{S5. Electrical characterization}
\begin{figure*}[ht]
\begin{center}
\includegraphics[width=5.5in]{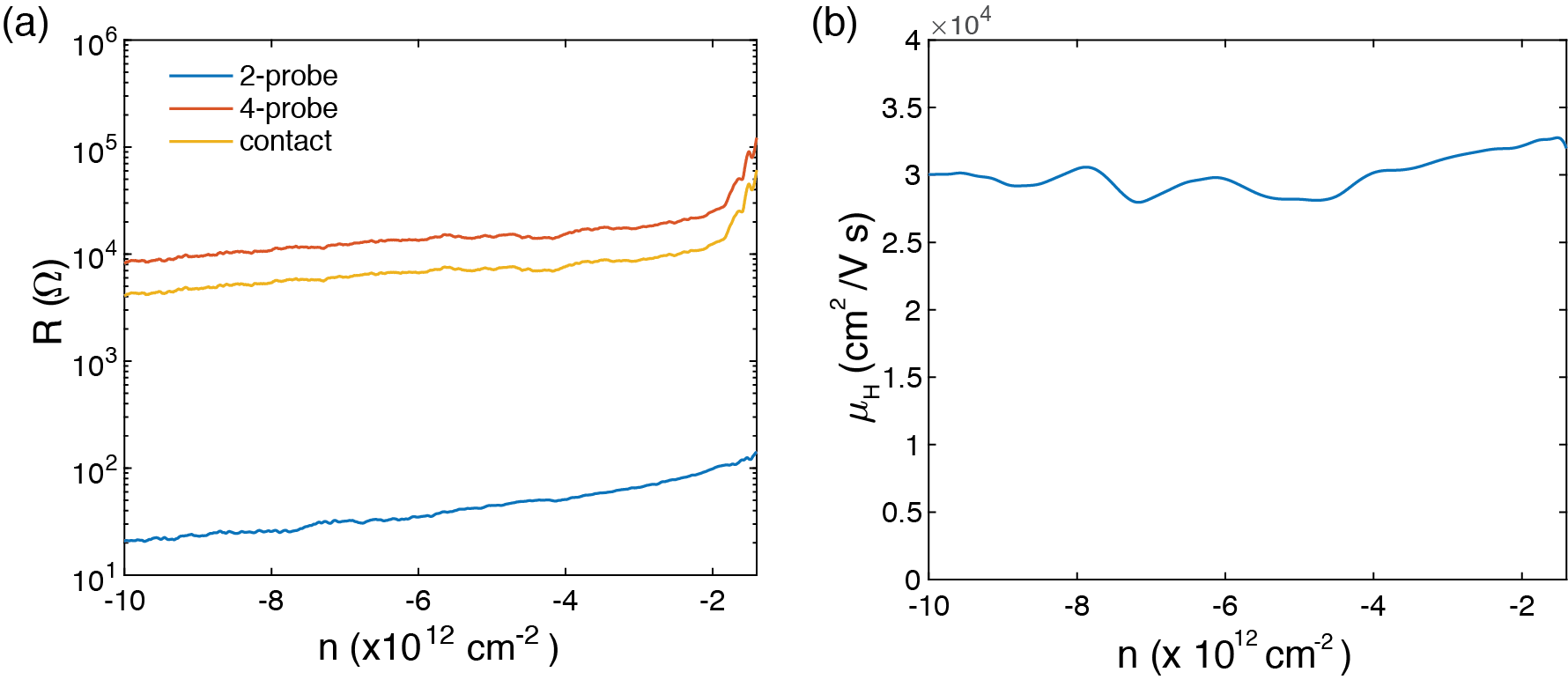}
\vspace{-0.15 in}
\end{center}
\caption{
(a) Comparison between 2-probe resistance $R_{2p}$, 4-probe resistance $R_{4p}$ and contact resistance $R_c$ versus charge density. The contact resistance is derived by $R_c = R_{2p}-(L_{tot}/L_{in})R_{4p}$, where $L_{tot}$ and $L_{in}$ are the total and inner channel lengths, respectively, and $L_{tot}/L_{in} \approx 2$.
(b) Hall mobility versus charge density. $\mu_H = (L/W)dR_{xy}/dB/R_{xx}$, where L and W are the channel aspect ratio$\approx 1$, $dR_{xy}/dB$ is calculated at low field.
}
\label{electrical characterization}
\vspace{-0.15 in}
\end{figure*}

\end{widetext}
\end{document}